\documentstyle[bibnorm]{lamuphys}
\makeatletter
\let\chapter\hid@chapter
\makeatother
\begin{document}
\pagenumbering{arabic}

\title{Nonextensive statistical effects in nuclear physics problems}

\author{G. Kaniadakis\inst{1}, A. Lavagno\inst{1,2}, 
M. Lissia\inst{3,4} and P. Quarati\inst{1,3}
}

\institute{
Dipartimento di Fisica, Politecnico di Torino, 
I-10129 Torino, Italy 
\and
Istituto Nazionale di Fisica Nucleare, Sezione di Torino, 
I-10125 Torino,
Italy 
\and
Istituto Nazionale di Fisica Nucleare, Sezione di Cagliari,
I-09042 Monserrato, Italy
\and
Dipartimento di Fisica, Universit\`a di Cagliari, 
I-09042, Monserrato, Italy }

\maketitle

\begin{abstract}
Recent progresses in statistical mechanics indicate the Tsallis
nonextensive thermostatistics as the natural generalization of the
standard classical and quantum statistics, when memory effects and
long range forces are not negligible. In this framework,
weakly nonextensive statistical deviations can strongly reduce
the puzzling discrepancies between experimental data and theoretical
previsions for solar neutrinos and for pion transverse-momentum
correlations in Pb-Pb high-energy nuclear collisions.
\end{abstract}

\section{Introduction}

Nuclear and, in general, many-particle systems are
often in regimes where concepts of statistical 
physics come into play. Long-range forces, 
memory effects, sizeable correlations and fluctuations of 
the observables provide demanding testing grounds of the 
equilibrium and non-equilibrium statistical mechanics theory.  

In many physical applications, the kinetic description of 
the system is developed in the weak-coupling limit defined 
by the condition $\tau_b\ll (\tau_\lambda$, $\tau_u)$, 
where $\tau_b$, $\tau_\lambda$ and $\tau_u$ are the duration 
of a binary collision, the mean free time between 
subsequent collisions and the characteristic time of the 
mean-field potential, respectively \cite{abe}. 
In this case, the many-body collision process is well described 
independent binary collisions in the framework of 
the standard Boltzmann equation and in terms of the $\delta(t-t^{'})$ 
function for the fluctuating stochastic Langevin force 
(Markovian approximation). 

If the mean-field characteristic time becomes comparable 
to the duration of a binary collision ($\tau_b\le\tau_u$), 
the independent binary collision approximation is no longer 
correct and the Markovian description breaks down: memory 
effects become important.  

Tsallis \cite{tsa} has recently advanced a generalization of the
conventional Boltzmann-Gibbs thermostatistics that
overcomes the inability of the conventional statistical mechanics 
to tackle those many physical problems with long-range interactions,
long-range microscopic memory, or fractal space-time constraints. 
There exist already many applications:
astrophysical self-gravitating systems  \cite{plasg}, 
the solar neutrino problem \cite{qua}, distribution of peculiar 
velocities of galaxy clusters \cite{lava}, cosmology \cite{placm}, 
many-body theory, dynamical linear response theory 
and variational methods \cite{raja2}.

In this paper, after a brief review of the Tsallis thermostatistics 
in Sec. II, we discuss (Sec. III) how a weakly non-ideal stellar 
plasma, such as the solar interior, can produce an equilibrium 
velocity distribution that deviates from the standard one and how 
this effect can be relevant to the solar neutrino problem.
In Sec. IV, we consider the interpretation of 
the pion transverse-momentum NA49 experimental data in high-energy 
Pb-Pb collisions in the framework of the nonextensive thermostatistics. 
Although the physics of the two problems is very different, 
they are both characterized by memory effects and 
long-range forces and, consequently, show nonextensive statistical behaviors. 

\section{Nonextensive statistical mechanics}

The principal features of the Tsallis generalized thermostatistics is
based upon the following two postulates \cite{tsa}. 

\begin{itemize}
\item Given $p_k$ the probability in any $W$ different microstates 
$k$, the entropy of a system is defined as 
\begin{equation}
S_q=\frac{1}{q-1}\, \sum_{k=1}^W p_k \, (1-p_k^{q-1}) \;,
\label{tsaen}
\end{equation}
where $q$ is a fixed real parameter. The generalized entropy has 
the usual properties of positivity, equiprobability, concavity, 
irreversibility and, in the limit $q\rightarrow 1$, is equal
to the conventional Boltzmann-Gibbs entropy $S=-\sum_k p_k \log p_k$.

\item The mean value of an observable ${\cal O}$, whose
value in the microstate $k$ is ${\cal O}_k$, is defined as  
\begin{equation}
\langle {\cal O}\rangle_q=\sum_{k=1}^{W} p_k^q \, {\cal O}_k \;.
\label{mval}
\end{equation}

\end{itemize}

The $S_q$ entropy is nonextensive. If $A$ e $B$ are two 
independent systems $A$ e $B$, {\em i.e.}, the probability of composite
system $A+B$  factorizes into $p_{A+B}(u_A, u_B)=p_A(u_A) \, p_B(u_B)$),
the global entropy is not the sum of the entropies of the subsystems, but
\begin{equation}
S_q(A+B)=S_q(A)+S_q(B)+(1-q) S_q(A)S_q(B) \; .
\end{equation}

The single particle distribution function is obtained by the usual procedure
of maximizing the Tsallis entropy with the constraints that the average
internal energy and the average number of particles remain constant:
\begin{equation}
f(v)=\left[1-(1-q) \frac{m v^2}{2 kT}\right]^{1/(1-q)}\;.
\label{tsadi}
\end{equation}
When the entropic parameter $q$ is smaller than $1$ 
the distribution has an upper cut-off:
$m v^2/2 \leq kT/(1-q)$ (the tail is depleted). 
The distribution correctly
reduces to the exponential Maxwell-Boltzmann distribution in the limit
$q\to 1$.
When the parameter $q$ is greater than 1, there is no cut-off
and the (power-law) decay is slower than exponential
(the tail is enhanced).

These classical distribution can be generalized 
to the quantum case under some not too restrictive approximation 
(see Ref.\cite{buyu} for details) resulting in the following analytical
expression for the mean occupational number:
\begin{equation}
\langle n\rangle_q=\frac{1}{[1+(q-1)\beta (E-\mu)]^{1/(q-1)}\pm 1} \;,
\label{distri}
\end{equation}
where $\beta=1/kT$, and the $+$ ($-$) sign applies to fermions (bosons).
In the limit 
$q\rightarrow 1$ (extensive statistics), we recover the conventional 
Fermi-Dirac and Bose-Einstein distribution.

\section{Non-ideal stellar plasma and solar neutrino fluxes}

\subsection{Plasma parameter and solar interior}

The stellar core is usually described as an ideal plasma in the
Debye-H\"uckel mean-field approximation. However, density 
and temperature conditions in stellar plasmas, such as the core of the
Sun, of brown dwarfs or of Jupiter, suggest that the 
microscopic diffusion of electrons and ions could be nonstandard 
and that the Debye-H\"uckel approximation is not sufficiently accurate.

The appropriate theoretical approach and the effective interactions 
used to describe a plasma can be deduced from its plasma parameter $\Gamma$ 

\begin{equation}
\Gamma=\frac{(Z e)^2}{a \, kT}  \ \ ,
\end{equation}
where $a = n^{-1/3}$ is of the order of the interparticle average distance
($n$ is the average density). The plasma parameter is a measure of the ratio
of the mean (Coulomb) potential energy and the mean kinetic (thermal) energy.
On the basis of its value, we can distinguish three different regimes. 
\begin{itemize}
\item 
$\Gamma \ll 1$. The plasma is described by the Debye-H\"uckel
mean-field theory as a dilute weakly-interacting gas.
The screening Debye length 
\begin{equation}
R_D=\sqrt{ \frac{kT}{4\pi e^2 \sum_i Z^2_i n_i} } \ \ ,
\end{equation}
is much greater than the average interparticle distance $a$, hence there
is a  large number of particles in the Debye sphere
($N_D\equiv(4\pi/3)R_D^3$).
Collective degrees of freedom are present (plasma
waves), but they are weakly coupled to the individual degrees of freedom
(ions and electrons) and, therefore, do not affect their distribution.
Binary collisions through screened forces produce
the standard velocity distribution.

\item 
$\Gamma \approx 0.1$. The mean Coulomb energy potential is not much
smaller than the thermal kinetic energy and the screening length 
$R_D\approx a$. It is not possible to clearly separate individual and
collective degrees of freedom. The presence of at least two different
scales of energies of the same rough size produces deviations from the
standard statistics which describe the system in terms of a single
scale, $kT$.

\item 
$\Gamma > 1$. This is a high-den\-si\-ty/low-tem\-pe\-ra\-tu\-re plasma; the
Coulomb interaction and quantum effects start to dominate and determine the
structure of the system.
\end{itemize}

In the solar interior the plasma parameter is
$\Gamma_\odot\approx 0.1$; therefore, the solar core is a weakly non-ideal
plasma where the Debye-H\"uckel conditions are only
approximately verified. 

The reaction time necessary to build up screening after a hard 
collision can
be estimated from the inverse solar plasma frequency 
$t_{pl}=\omega_{pl}^{-1} = 
\sqrt{m/ (4\pi n e^2 )} \approx 10^{-17}$~sec, 
and it is comparable to the
collision time $t_{coll} = \langle\sigma v n \rangle^{-1} \approx 10^{-17}$ 
~sec. Therefore, several collisions are likely necessary before the particle
looses memory of the initial state and the scattering
process can not be considered Markovian. 
In ad\-di\-tion to many-body col\-li\-sio\-nal effects, electric 
microfields are present and the distribution and the fluctuations 
of these microfields must be carefully considered, since they modify
the usual Boltzmann kinetics. 

At the light of the above considerations, we conclude that density 
and temperature conditions of the solar core suggest that non-Markovian 
memory effects and long-range forces are present. Because the solar core 
lies in a intermediate region of the plasma parameter, the corrections 
to the standard Maxwell-Boltzmann distribution should be small and affect 
mainly the more energetic particles. However we will see in next 
subsection that this small deviation strongly affects the nuclear 
reaction rates and, consequently, the determination of the solar 
neutrino fluxes.

\subsection{Thermonuclear reaction rate and solar neutrino fluxes}

The nuclear reaction rates in the stellar interiors play a 
crucial r\^{o}le in the understanding of the structure and evolution of stars. 

The reaction rate per particle pair is defined as
\begin{equation}
\langle v \sigma \rangle = \int_0^\infty \! f(v)\,  \sigma v \, dv \, ,
\label{sigmav}
\end{equation}
where the particles distribution function $f(v)$ is a local function
of the temperature.

The kinetic energy of the ions in the solar core is essentially the 
thermal energy $kT_\odot\approx 1.36$ keV which is far below the 
Coulomb barrier, therefore only a small
number of particles in the high-energy tail of the distribution has a
chance of reacting: 
this high-energy tail plays a crucial r\^{o}le for the reaction rates,
which becomes very sensitive to small changes of the distribution.

For small deviations from the standard extensive equilibrium 
statistics, we can write the generalized Tsallis distribution 
(\ref{tsadi}) as follows \cite{qua}
\begin{equation}
f(E) \sim (kT)^{-3/2} \,  e^{-E/kT - \delta (E/kT)^2 } \, ,
\end{equation}
where $\delta=(1-q)/2$.

The presence of even a tiny deviation from MB in the solar core produces
large changes of the subbarrier nuclear reaction rates and, consequently,
of the predicted neutrino fluxes. Following the general homology 
relationships for the variations of physical inputs,
see for instance Ref.~\cite{castell97}, we can estimate the effect of
the non-Maxwel\-lian distribution on the fluxes~\cite{qua}:
\begin{equation}
R_j=\frac{\Phi_j}{\Phi_j^{(0)}}=e^{-\delta_j \beta_j} \, ,
\label{flussi}
\end{equation}
for the fluxes $j=$ $^7$Be, $^8$B, $^{13}$N and $^{15}$O, while we
use the solar luminosity constraint to determine the $pp$ flux, 
$R_{pp} = 1+0.087\times(1-R_{Be})
                       +0.010\times(1-R_{N})
                       +0.009\times(1-R_{O})$, and keep fixed
the ratio $\xi\equiv\Phi_{pep}/\Phi_{pp}=2.36\times 10^{-3}$. 
The power indices $\beta_j$, that appear in Eq.~(\ref{flussi}), depend
on the nuclear reaction considered and their values have been taken from
Ref.~\cite{castell97}. 
In principle, there could be a different parameter $\delta_j$ for
each reaction and it should be possible to calculated them from the 
specific interactions in the solar plasma core. For the purpose of
estimating the possible effects of this mechanism on the solar neutrino
fluxes, we used two simple models where $\delta_j$ were used as free
parameters. 

The first model uses the same $\delta$ for all the
reactions; the best fit to the experimental data give $\delta=0.005$
with a corresponding $\chi^2=35$. The second model fits
two different $\delta$'s, one ($\delta_{(17)}$) for the 
$p+{}^7$Be reaction and the other ($\delta_{(34)}$) for the
${}^3$He + ${}^4$He reaction. The best result gives $\chi^2=20$ with
$\delta_{(17)}=-0.018$ and $\delta_{(34)}=0.030$.
The negative value of $\delta_{(17)}$ means that the corresponding
distribution has an enhanced tail and that the $p+{}^7$Be reaction
rate increases. 

In spite of the fact that the values of $\chi^2$ are much smaller than
the ones in the SSM ($\chi^2_{SSM}>74$), they are still large: this
mechanism cannot solve the solar neutrino problem. However, even if
$\delta$ is small, it has non-trivial consequences on the neutrino fluxes:
the boron (beryllium) flux can change of as much as 50\% (30\%).

\section{Fluctuations and correlations in high-energy nuclear collisions}

The recent event-by-event analysis of central Pb+Pb collisions at 158 A GeV 
(NA49 coll.) has spurred great interest because of the strong suppression 
of the pion transverse-momentum fluctuations in Pb+Pb respect to p+p 
collisions. A reduction of the fluctuations appears reasonable because 
correlations should be washed out by meson-baryon and meson-meson
rescatterings; however, the size of this suppression is not understood. 

Ga\'zdzicki and Mr\'owczy\'nski \cite{gaz92} have introduced a definition
of the event-by-event transverse-momentum fluctuations that is independent
of the particle multiplicity:
\begin{equation}
\Phi_{p_\perp}=\sqrt{\langle Z_{p_\perp}^2 \rangle \over \langle N \rangle} -
\sqrt{\overline{z_{p_\perp}^2}} \;,
\label{phix}
\end{equation}
where $z_{p_\perp} = p_\perp - \overline{{p_\perp}}\,$ is a single-particle 
measure of the deviation of ${p_\perp}$ from its average value, and where
$Z_{p_\perp} = \sum_{i=1}^{N}({p_\perp}_i - \overline{{p_\perp}})$ is the
corresponding total contribution ($N$ is the number of the particles in the
event).

In a very recent paper, Mr\'owczy\'nski \cite{mro} has calculated this
correlation measure $\Phi_{p_\perp}$ for a pion gas in global equilibrium
within the standard extensive thermodynamics. His result
($\Phi_{p_\perp}=6.5$ MeV at $T=120$ MeV) is sensibly greater 
than the experimental value $\Phi_{p_\perp}=0.7\pm 0.5$ MeV, which has
been measured in the central Pb+Pb collision by the NA49
collaboration \cite{rol}). 

What is the origin of such a large discrepancy between theoretical results
and experimental data? 
Can the conditions of density and temperature at the early 
stage collisions modify the correlations between the produced particles?

The quoted theoretical result \cite{mro} has been obtained within
the framework of the conventional quantum statistical mechanics. 
Therefore, pions, which are sufficiently light to show quantum degeneracy, 
follow the standard Bose-Einstein distribution.
However, it is common opinion that, because of the extreme conditions of 
density and temperature in ultrarelativistic heavy ion collisions, 
memory effects and long-range color interactions affect the thermalization
process \cite{gav,hei,biro}. In fact, if the quark-gluon plasma forms
at the early stage of the collision, strong chromoelectric fields appear
within the parton gas. Recent investigations, both for for QED and QCD,
have shown that these strong fields are connected to the presence of
non-Markovian process in the kinetic equation \cite{rau}. 

The foregoing considerations suggest to calculate the equilibrium correlation
measure in the framework of the nonextesive statistics.

On the basis of the generalized thermodynamics relations \cite{tsa}, 
it can be shown that 
\begin{eqnarray}
\overline{z^2_{p_{\perp}}} =
{1 \over \rho}\int{d^3p \over (2\pi )^3} \, \, \Big(p_{\perp} - \overline{p}_{\perp} \Big)^2 \; \langle n\rangle_q \; ,
\label{qz2p}
\end{eqnarray}
where $\langle n\rangle_q$  is the mean occupation number of 
Eq.(\ref{distri}) and 
\begin{eqnarray}
{\langle Z_{p_{\perp}}^2 \rangle \over \langle N \rangle }=
{1 \over \rho}\int{d^3p \over (2\pi )^3}
\,\Big(p_{\perp} - \overline{p}_{\perp} \Big)^2 \,
\langle\Delta n^2\rangle_q \;,
\label{qz2g}
\end{eqnarray}
where 
\begin{equation}
\overline{p}_{\perp} = {1 \over \rho}\int{d^3p \over (2\pi )^3} \;
p_{\perp} \; \langle n\rangle_q \;, \ \ \ \ \ \ 
\rho = \int{d^3p \over (2\pi )^3} \;\langle n\rangle_q 
\label{rho}
\end{equation}
and 
\begin{equation}
\langle\Delta n^2\rangle_q\equiv
\frac{1}{\beta}\frac{\partial\langle n\rangle_q}{\partial\mu}=
\frac{\langle n\rangle_q }{1+(q-1)\beta (E-\mu)}\, 
(1\mp \langle n\rangle_q)\; 
\label{fluc}
\end{equation}
are the particle fluctuations in nonextensive ($q\ne 1$) statistics.  
For $q=1$, we recover the well known fluctuations expression for fermions
($-$) and bosons ($+$). 
From Eq.~(\ref{fluc}), we see that the expression for the generalized
fluctuations has the same structure of the standard one
modified by the factor $1/[1+(1-q) \beta (E-\mu)]$.

For the boson case a value $q>1$ implies a very strong suppression of  
$\Phi_{p_\perp}$. If we consider a pion gas and fix the freeze-out
temperature to $T=120$ MeV (obtained from the analysis of single particle
spectra), we reproduce the experimental value ($\Phi_{p_\perp}=0.7\pm 0.5$
MeV) using $q=1.015$. Hence, a small deviation from the standard statistics
($q-1=0.015$) is sufficient for eliminating the puzzling discrepancy
between theoretical calculations and experimental data. 

Finally, we observe that fluctuations are the more strongly modified
the larger the mass of the detected particle for a given
value of the deformation parameter $q$. 
Hence, this model predicts  stronger nonextensive effects for correlations
of heavier mesons and baryons: future measurements of $\Phi_{p_\perp}$ for
these particles should find an even larger reduction when plasma is
formed. New data and investigations are necessary to gain a deeper 
understanding of the high-energy heavy-ion observables.

\section{Conclusion}
We have considered the relevance of the generalized nonextensive statistics
to the the solar neutrino problem and to the interpretation of the 
pion transverse-momentum correlations in Pb-Pb high energy collisions 
experimental data. 
Although the two problems involve rather different physics, both
appear to show memory effects and long range forces. Discrepancies
between experimental data and theoretical models can be strongly reduced
when weakly-nonextensive ($|q-1| < 0.02$) statistical effects are considered.

In the solar neutrino problem a small deviation from the 
Maxwell-Boltz\-mann distribution produces strong modifications
of the thermonuclear reaction rates and, consequently, modifies
neutrino fluxes by amounts comparable to those that constitute the 
solar neutrino problem without affecting bulk properties such as 
the sound speed or hydrostatic equilibrium. 

In high-energy ion collisions, nonextensive statistics
predicts a reduction of the correlation measure $\Phi_{p_\perp}$
in Pb-Pb collisions ($\Phi_{p_\perp}=0.7\pm 0.5$ MeV)
respect to p-p collisions ($\Phi_{p_\perp}=4.2\pm 0.5$ MeV)
due to plasma effects.
This reduction has been seen in the experiments and could be 
interpreted as a signature of the transition to a quark-gluon-plasma
phase driven by the extreme conditions of density and temperature in the
early stage of the collision.

\end{document}